# MINING DEVELOPER COMMUNICATION DATA STREAMS

Dr Andy M. Connor[1], Dr Jacqui Finlay[2] and Dr Russel Pears[2]

[1] CoLab, Auckland University of Technology, Private Bag 92006, Wellesley Street, Auckland, NZ
`andrew.connor@aut.ac.nz`

[2] School of Computing & Mathematical Sciences, Auckland University of Technology, Private Bag 92006, Wellesley Street, Auckland, NZ

*ABSTRACT*

*This paper explores the concepts of modelling a software development project as a process that results in the creation of a continuous stream of data. In terms of the Jazz repository used in this research, one aspect of that stream of data would be developer communication. Such data can be used to create an evolving social network characterized by a range of metrics. This paper presents the application of data stream mining techniques to identify the most useful metrics for predicting build outcomes. Results are presented from applying the Hoeffding Tree classification method used in conjunction with the Adaptive Sliding Window (ADWIN) method for detecting concept drift. The results indicate that only a small number of the available metrics considered have any significance for predicting the outcome of a build.*

*KEYWORDS*

*Data Mining, Data Stream Mining, Hoeffding Tree, Adaptive Sliding Window, Jazz*

## 1. INTRODUCTION

Software development projects involve the use of a wide range of tools to produce a software artefact, and as a result the history of any given software development may be distributed across a number of such tools. Recent research in this area [1] has described the different types of artefacts that can be used to reconstruct the history of a software project. These include the source code itself, source code management systems, issue tracking systems, messages between developers and users, meta-data about the projects and usage data.

IBM's Rational Team Concert is a fully integrated software team collaboration and development tool that automatically captures software development processes and artefacts [2]. The tool has been produced as part of the Jazz project at IBM and the development repository has been released for research purposes. The Jazz repository contains real-time evidence that provides researchers the potential to gain insights into team collaboration and development activities within software engineering projects [3]. With Jazz it is possible to extract the interactions between contributors in a development project and examine the artefacts produced. Such interactions are captured from user comments on work items, which is the primary communication channel used within the Jazz project. As a result Jazz provides the capability to extract social network data and link such data to the software project outcomes.

This paper describes an attempt to fully extract the richness available in the IBM Jazz data set by representing the emergence of develop communication as a data stream as a means of predicting software build outcomes. Traditional data mining methods and software measurement studies are tailored to static data environments. These methods are typically not suitable for streaming data which is a feature of many real-world applications. Software project data is produced continuously and is accumulated over long periods of time for large systems. The dynamic nature of software and the resulting changes in software development strategies over time causes changes in the patterns that govern software project outcomes. This phenomenon has been recognized in many other domains and is referred to as data evolution, dynamic streaming or concept drifting. However there has been little research to date that investigates concept drifting in software development data. Changes in a data stream can evolve slowly or quickly and the rates of change can be queried within stream-based tools. This paper describes an initial attempt to fully extract the richness available in the Jazz data set by constructing predictive models to classify a given build as being either successful or not, using developer communication metrics as the predictors for a build outcome.

## 2. BACKGROUND & RELATED WORK

The mining of software repositories involves the extraction of both basic and value-added information from existing software repositories [4]. The repositories are generally mined to extract facts by different stakeholders for different purposes. Data mining is becoming increasingly deployed in software engineering environments [3, 5, 6] and the applications of mining software repositories include areas as diverse as the development of fault prediction models [7], impact analysis [8], effort prediction [9, 10], similarity analysis [11] and the prediction of architectural change [12] to name but a few.

According to Herzig & Zeller [2], Jazz offers huge opportunities for software repository mining but such usage also comes with a number of challenges. One of the main advantages of Jazz is the provision of a detailed dataset in which all of the software artefacts are linked to each other in single repository. This simplifies the process of linking artefacts that exist in different repositories. To date, much of the work that utilizes Jazz as a repository has focused on the convenience provided by this linking of artefacts, such as bug reports to specification items, along with the team communication history. Researchers in this area have focused on areas such as whether there is an association between team communication and build failure [13] or software quality in general [14]. Other work has focused on whether it is possible to identify relationships among requirements, people and software defects [15].

Previous research has investigated the prediction of build outcomes for the Jazz repository by developing decision models based on the extraction of software metrics from the source code included in the repository [16] including the modeling of the available data as a data stream in order to apply data stream mining techniques [17] to facilitate the mining of project data on the fly to provide rapid just-in-time information to guide software decisions. The research presented in this paper extends the work of applying data stream mining techniques by considering the role of developer communication metrics in the prediction of build outcomes. The data used in this work is consistent with previous work [17] in that the developer communication metrics are extracted from the same builds as were used for extracting software metrics which allows direct comparison of the predictive power of the two approaches.

## 3. THE JAZZ REPOSITORY

What makes the Jazz repository unique is that there is full traceability between a wide range of software artefacts. The Jazz team itself is globally distributed and therefore the existing repository data provides the opportunity to data mine developer communication, bug databases and version control systems at the same time. Within the repository each software build may have a number of work items associated to it. These work items represent various types of units of work and can represent defects, enhancements and general development tasks. Work items provide traceable evidence for coordination between people as they can also be commented. In addition to this they are one of the main channels of communication and collaboration used by contributors of the Jazz project.

That being said there are, of course, other channels of communications which are not captured by work items, these include email, on-line chats and face-to-face meetings. Even though these elements are not captured, exploration of communication on work items offers a non-intrusive means to explore much of the collaboration that has occurred during the Jazz project. The Jazz team itself is fairly large, with 66 team areas for approximately 160 contributors that are globally distributed over various sites across the United States of America, Canada and Europe.

To explore the communication between contributors involved in builds, social network metrics are derived from the communication networks that are present within work items. Each work items is able to be commented on and this is the main task-related communication channel for the Jazz project. This enables contributors to coordinate with each other during the implementation of a work item. There are many elements, in regards to contributors, of a work item to consider. This makes the process of constructing a social network a little more challenging. In doing so some basic assumptions about the data has been made. Work items can have various contributors assigned to various roles, for example there are creators, modifiers, owners, resolvers, approvers, commenters and subscribers. For the purposes of this research a social network is constructed similar to the work presented by Wolf et al. [13].

For each social network constructed, nodes represent contributors involved with a build. A series of directed edges/links represent the communication flow from one contributor to another. A build can have any number of work items associated with it. Therefore the social networks generated at the work item level are required to be propagated to the build level for analysis of its impact on build success. To do this if a contributor appears within multiple work items that are associated with a single build, only one node is created to represent that contributor (there are no duplicate nodes). However additional edges are added to reflect entirely new instances communication that takes place between contributors. All edges within a network are treated as unique (there are no duplicate edges). This is because it would threaten the validity of metrics such as density. For example if a as a network that is fully connected has a density of 1. If there are edges which represent each individual flow of communication the density metric would no longer be valid (potentially being greater than 1), which would make comparisons between networks metrics challenging.

For this research, roles which are used to construct the network nodes include committers of change sets, creators, commenters and subscribers. Committers of change sets for a build are presented as node, as they have a direct influence on the result of the build. Creators of a work item are communicating the work item itself with other members of the team. Commenters are contributors that are discussing issues about a work item. Subscribers are people who may be interested on the status of a work item as it has impact on their own work/other modules. In order to generate the edges between nodes, rules have been implemented to establish connections between people.

From these elements constructing the social networks for each build, the metrics are calculated are:

- Social Network Centrality Metrics:
    - Group In-Degree Centrality, Group Out Degree Centrality, Group InOut-Degree Centrality, Highest In-Degree Centrality, Highest Out-Degree Centrality
    - Node Group Betweenness Centrality and Edge Group Betweenness Centrality
    - Group Markov Centrality

- Structural Hole Metrics:
  - Effective Size and Efficiency

- Basic Network Metrics:
  - Density, Sum of vertices and sum of edges

- Additional Basic Count Metrics:
  - Number of work items the communication metrics were extracted from
  - Number of change sets associated with those work items

The process of generating the underlying data consisted of first selecting the appropriate builds for analysis. As the intention was to allow comparison with previous work [17], only builds that had associated source code were selected. In total this resulted in 199 builds of which 127 builds were successful and 72 failed. The builds were comprised of 15 nightly builds (incorporating changes from the local site), 34 integration builds (integrating components from remote sites), 143 continuous builds (regular user builds) and 7 connector Jazz builds. Builds were ordered chronologically to simulate the emergence of a stream of data over time and then work items extracted for each build to allow the social network for the build to be constructed.

## 4. DATA STREAM MINING

Most software repositories are structured that over time the underlying database grows and evolves resulting in large volumes of data. The data from this perspective arrives in the form of streams. Data streams are generated continuously and are often time-based. In large and complex systems the data, arriving in a stream form, takes its toll on resources, particular the need for large capacity data storage. In some cases it is impossible to store the entire steam, which is the case for the Jazz repository. This is because streams themselves can be overwhelming. Often in these cases the data is processed once and then is disposed of.

The implications of data stream mining in the context of real-time software artefacts is yet to be fully explored. Currently there is little research that has explored whether or not stream mining methods can be used in Software Engineering in general, let alone for predicting software build outcomes. In large development teams software builds are performed in a local and general sense. In a local sense developers perform personal builds of their code. In a general sense the entire system is built (continuous and integration builds). These builds occur regularly within the software development lifecycle. As there can be a large amount of source code from build to build, the data and information associated with a build is usually discarded due to system size constraints. More specifically, in the IBM Jazz repository, whiles there are thousands of builds performed by developers; only the latest few hundred builds can be retrieved from the repository. Data stream mining offers a potential solution to provide developers real-time insights into fault detection, based from source code and communication metrics. In doing so it enables developers to mitigate risks of potential failure during system development and maintenance and track evolutions within source code over time.

This work revolves around the use of a data stream mining techniques for the analysis of developer communication metrics derived from the IBM Jazz repository. For this purpose the Massive Online Analysis (MOA) software environment was used [29]. Data streams provide unique opportunities as software development dynamics can be examined and captured through the incremental construction of models over time that predict project outcome. Two outcomes are possible: success, or failure. A successful outcome signals that each of the constituent work items in a project has been built as per the specification and that the items have been integrated with each other without generating any errors. On the other hand, the failure outcome is caused by one or more work items malfunctioning and/or errors being generated in the integration process.

Instances within the stream are sorted via the starting date/time property of a software build (oldest to newest) to simulate software project build processes. Using the Hoeffding tree [18], a model is built using the first 20 instances for training. Prediction is performed on each of the remaining 179 instances which are used to incrementally update the model built. Furthermore, as the outcomes are known in advance, model accuracy can be evaluated on an ongoing basis. This was possible as we used a static dataset which contained pre-assigned class labels to simulate a data stream environment. The methods used to mine the simulated data stream are described in the following section.

### 4.1. Hoeffding Tree

Decision trees were selected as the machine learning outcome for this research. This choice was influenced by the fact that the decision tree has proven to be amongst the most accurate of machine learning algorithms while providing models with a high degree of interpretability. However, the basic version of the decision tree algorithm in its basic form cannot be deployed in a data stream environment as it is incapable of incrementally updating its model, a key requirement in a data stream environment. As such, an incremental version, called the Hoeffding tree was deployed.

The Hoeffding tree is a commonly used incremental decision tree learner designed to operate in a data stream environment [19]. Unlike in a static data environment decision tree learners in a data stream environment are faced with the difficult choice of deciding whether a given decision node in the tree should be split or not. In making this decision, the information gain measure that drives this decision needs to take into account not just the instances accumulated in the node so far, but must also make an assessment of information gain that would result from future, unseen instances that could navigate down to the node in question from the root node. This issue is resolved through the use of the Hoeffding bound [20].

The Hoeffding bound is expressed as:

$$\epsilon = \sqrt{\frac{R^2 \ln\left(\frac{1}{\delta}\right)}{2n}}$$

Where R is a random variable of interest, n is the observations and d is a confidence parameter. Essentially, the Hoeffding bound states that with confidence 1-δ, the population mean of R is at least $\bar{r}$ - $\epsilon$, where $\bar{r}$ is the sample mean computed from n observations.

In the context of data mining, the variable R is the information gain measure which ranges from 0 to 1 in value. One key advantage of the bound is that it holds true irrespective of the underlying data distribution, thus making it more applicable than bounds that assume a certain distribution, for example the Gaussian distribution.

The Hoeffding tree implementation available from MOA was used and coupled with a concept drift mechanism called the Adaptive Sliding Window (ADWIN) which was also available from the MOA environment. Most machine learner algorithms in a data stream environment use fixed size sliding window widths to incrementally maintain models. Sliding windows offer the advantage of purging old data while retaining a manageable amount of recent data to update models. However, while the concept of windows is attractive in the sense that memory requirements are kept within manageable bounds, fixed sized window are often sub-optimal from the viewpoint of model accuracy. This is due to the fact that concept change or drift may not align with window boundaries. When changes occur within a window, all instance before the change point should be purged leaving the rest of the instances intact for updating the model built so far. The ADWIN approach has many merits in terms of detecting concept drifts in dynamic data streams.

## 4.2. Concept Drift Detection

ADWIN works on the principle of statistical hypothesis testing. It maintains a window consisting of all instances that have arrived since the last detected change point. In its most basic form, the arrival of each new instance causes ADWIN to split the current window into two sub-windows, left and right. The sample means of the data in the two sub-windows are compared under a null hypothesis $H0$ that the means across the sub-windows are not significantly different from each other. If $H0$ is rejected, concept drift is taken to have occurred and ADWIN shrinks the window to only include instances in the right sub-window, thus removing instances in the left window representing the "old" concept. Simultaneously, the Hoeffding tree is updated to remove sub-tree(s) representing the old or outdated concept.

## 5. EXPERIMENTAL RESULTS

The experimental approach used in this work involves simulating a data stream by stepping through historical data in a sequential manner. The aim is to track key performance aspects of the predictive model as a function of time as well as also quantifying the level of drift in the features used by the model that determine build outcomes over the progression of the data stream over time. This experimentation revealed that the model was robust to concept drift as the overall classification accuracy recorded a steady increase over time.

Due to the limited size of the data set the default value of grace period for the Hoeffding tree is lowered from the default 200 instances. At the beginning of this set of experiments various grace periods were trailed to see whether or not the beginning set of training instances had an effect on the final classification accuracy. The results indicated that if the grace period is set too high it will result in a loss of final accuracy, as the initial model built is over fitted to the data. In terms of results for sections 4.10.1 and 4.10.2 a grace period of 20 was found to generate the highest level of accuracy for the 199 instances. The split confidence is 0.05 and the tie threshold option is set to 0.1.

## 5.1. Hoeffding Tree Classification

The graph presented in Figure 1 presents the trend of overall classification accuracy for builds over time using the Hoeffding Tree method for the developer communication metrics. It is observed that after approximately 100 builds the prediction accuracy begins to stabilize and improve. This is to be expected because at the start of the training process insufficient instances exist, resulting in model under-fitting. The final overall prediction accuracy is approximately 63% which is only a nominal improvement from the earlier prediction accuracies which start around 52%. The overall trend shows that, as more instances are trained, the classification accuracy steadily improves but does not appears to have the same predictive power as source code metrics for which the same datastream mining approach produces models that have an overall accuracy of approximately 72% [17]. Figure 2 and Figure 3 show the sensitivity measures for actual successful and failed builds.

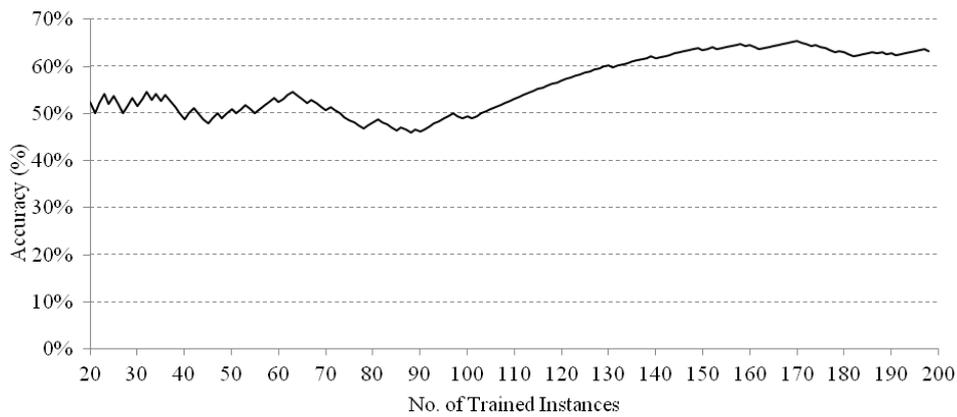

Figure 1. Overall prediction accuracy

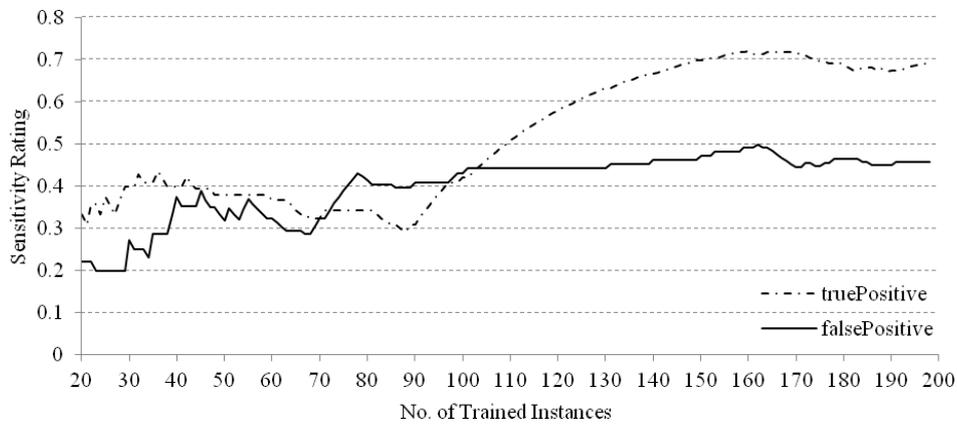

Figure 2. Sensitivity measures for successful builds

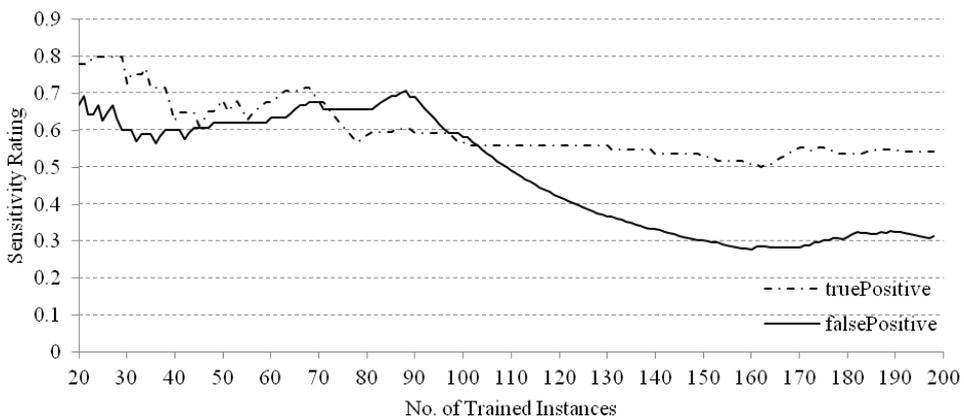

Figure 3. Sensitivity measures for failed builds

The results shown in Figure 2 and Figure 3 demonstrate that the evolution of the decision tree model is a complex scenario. Over the duration of the simulated data stream, the proportion of successful builds that are correctly classified by the decision tree steadily increases to around 70%., Over the same simulated period the proportion of failed builds that are correctly classified drops from a very high initial value to around 54%. Clearly the initial value is a result of the lack of data causing model under-fitting, but the outcomes of the end of simulating the data stream support observations made in previous work [16, 17] that it is significantly more challenging to identify a failed build than it is a successful one. In this case it appears that the developer communication metrics are equally as effective in this regard when compared to source code metrics [17]. From the false positive data it is clear that there is a significant problem in failed builds being misclassified as successful builds.

To fully understand the application of the Hoeffding tree approach it is important to analyze the emergence of the decision tree model, not just the final model itself. By examining Figure 1 it would seem reasonable to conclude that the minimum number of instances required to develop a classification tree that is reasonably stable would be around 100 instances. It is at this point that the prediction accuracy starts to stabilize and show a trend to improving asymptotically. However, an examination of the Hoeffding tree outcomes at this point shows that no actual decision tree has been generated by the model at this point in time. In fact, the Hoeffding tree approach has not identified a single feature that has sufficient predictive power to use effectively. The approach is therefore attempting to classify a new build in the data stream against the majority taken over all instances and all attribute values. So, for example, if there were 60% successful builds, then all builds would be labeled success. This is an exceptionally degenerate case where severe model under-fitting is occurring due to lack of training examples resulting in no clear predictors. The Hoeffding tree approach identifies an actual decision tree only after 160 builds. This first decision tree identifies only a single attribute against which to classify a given build and the resulting decision tree is shown in Figure 4.

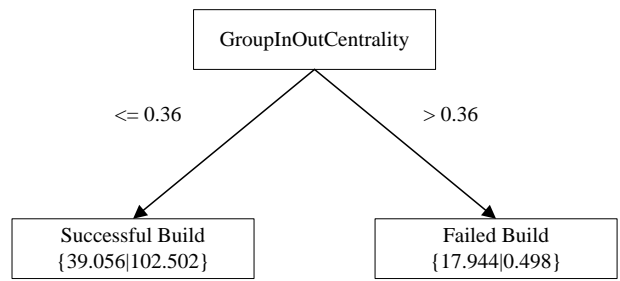

Figure 4. Decision tree at 160 builds

This decision tree indicates that the metric GroupInOutCentrality has emerged as the only significant predictor of success and failure. This outcome differs somewhat from previous work investigating the use of developer communication in predicting build outcomes for the Jazz repository [13] which indicated that there were no individual metrics that were statistical significant in terms of predicting outcomes. The leaves of the tree show the predicted outcome and the numeric values represent the votes used in the majority vote classifier. The value on left represents the weighted votes for failed builds and the value on the right represented the weighted votes for successful builds (i.e. failed builds | successful builds). Again, this indicates that there are a large number of failed builds being mis-classified as successful builds. So whilst it may appear that the outcomes differ from previous studies [13] there is not sufficient evidence to suggest that GroupInOutCentrality is a strong indicator of outcomes, particularly for the builds that would be of most interest, i.e. failed builds.

One of the advantages of the use of the ADWIN approach for concept drift detection and the result in terms of the Hoeffding tree approach, that results in only detecting and responding to changes that are statistically significant. After the initial model emerges at 160 builds there are no significant changes for a further 10 builds. At 180 builds such a change is detected but rather than see a different model appear it is expected that the model will evolve. The resulting model is shown in Figure 5.

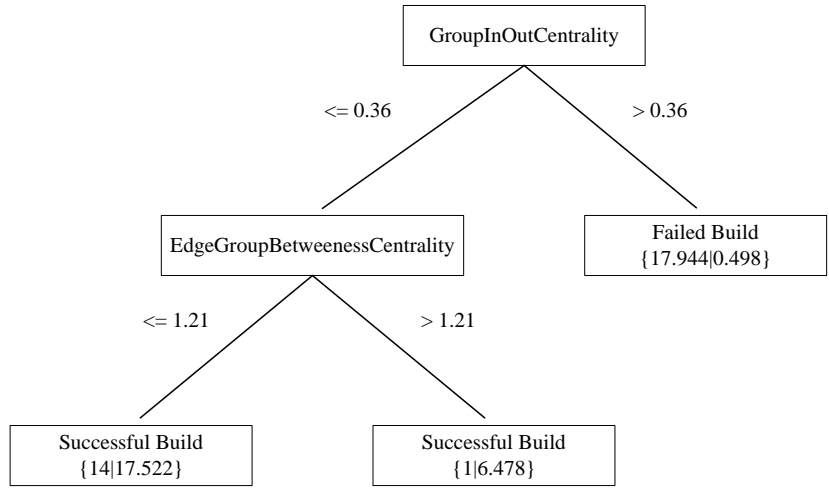

Figure 5. Decision tree at 180 builds

The addition of EdgeGroupBetweennessCentrality as a new metric has an interesting effect in that it refines the classification of successful builds only and does not improve the classification of failed builds at all. Of the 39 builds used in this incremental change, 15 of them are failed builds that are mis-classified as successful builds by the addition of the new metric. In percentage terms, the false positive rate for successful builds in these leaves of the tree is 38% which compares to a false positive rate for successful builds in Figure 4 of 27%. Therefore it is arguable that the model that has evolved is in fact a worse prediction model than its predecessor because whilst it has improved classification for successful builds this has been at the cost of a reduced ability to classify failed builds.

However, such variations are very small to the point of being insignificant and the reduction in quality of the model may be reversed in the long term as more data is captured.

### 5.2. Comparison with k-NN Clustering

A full validation of the Hoeffding tree approach by comparing to other methods has not yet been completed, however an initial comparison to the k-nearest neighbor (k-NN) algorithm has been undertaken. In this case, the initial 20 builds are excluded from the analysis and the k-NN algorithm is trained on the remaining 179 builds, of which 116 were successful builds and 63 were failed. The final prediction accuracies of the Hoeffding tree and the k-NN when applied to communication network metrics are shown in Table 1, where k = 5. The Hoeffding tree has performed better in terms of overall correctly classified instances for both successful and failed build outcomes when compared to the k-NN method.

Table 1. Comparison of Hoeffding Tree & k-NN (179 instances).

| | | Hoeffding Tree | k-NN |
|---|---|---|---|
| **Successful Builds** | *Correctly Classified Instances* | 83 | 80 |
| | *Incorrectly Classified Instances* | 33 | 36 |
| **Failed Builds** | *Correctly Classified Instances* | 32 | 23 |
| | *Incorrectly Classified Instances* | 31 | 40 |
| **Overall Accuracy of Prediction** | | 64.24% | 57.54% |

## 6. LIMITATIONS & FUTURE WORK

Most of the limitations in the current study are products of the relatively small sample size of build data from the Jazz project. With only 199 builds available it is difficult to truly identify significant metrics and evaluate the efficacy of the data stream mining approach. It has been observed that predicting failure is more challenging than predicting success and that not predicting failure doesn't mean that success has been predicted. This is due to the fact that the build successes and failures overlap in feature space and "failure" signatures have a greater degree of fragmentation than their "success" counterparts. This overlap is a strong symptom of the fact that some vital predictors of software build failure have not been captured in the Jazz repository. This is consistent with previous that utilized source code metrics for build outcome prediction. It is an open question as to whether any such predictors can indeed be quantified in a form suitable for use in a machine learning predictive context, however future work will investigate the combination of software metrics with developer communication metrics.

## 7. CONCLUSIONS

This paper presents the outcomes of an attempt to predict build success and/or failure for a software product by modeling the emergence of developer communication as a datastream and applying datastream mining techniques. Overall prediction accuracies of 63% have been achieved through the use of the Hoeffding Tree algorithm. This is a lower prediction accuracy than is obtained when source code metrics are mined as a datastream [17].

This research has presented a potential solution for encoding developer communication metrics as data streams. In the case of Jazz the data streams would be provided when a software build was executed, though this study simulated such a datastream from historical data. The real-time streams can be run against the model which has been generated from software build histories. From the real-time based predictions developers may delay a build to proactively make changes on a failed build prediction. One of the advantages of building predictive models using data stream mining methods is that they do not have large permanent storage requirements.

The results have shown that data stream mining techniques holds much potential as the Jazz environment, as the platform can continue to store the latest builds without losing relevant information for a prediction model that has been built over an extended series of (older) software builds. Future work will investigate the combination of developer communication metrics with source code metrics [17] into a unified predictive model.

## ACKNOWLEDGEMENTS

Our thanks go to IBM for providing access to the Jazz repository and the BuildIT consortium that has funded this research.

## Authors

Andy Connor is a Senior Lecturer in CoLab and has previously worked in the School of Computing & Mathematical Sciences at AUT. Prior to this he worked as a Senior Consultant for the INBIS Group on a wide range of systems engineering projects. He has also worked as a software development engineer and held postdoctoral research positions at Engineering Design Centres at the University of Cambridge and the University of Bath.

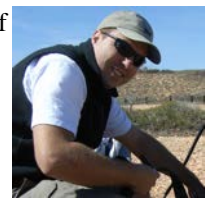

Jacqui Finlay completed her PhD in Computer Science at Auckland University of Technology, investigating the use of data mining in the use of software engineering management. Previously she was employed as Senior Systems Architect at GBR Research ltd.

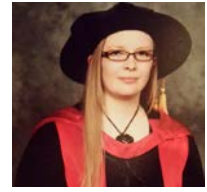

Russel Pears is a Senior Lecturer in the School of Computing & Mathematical Sciences at AUT, where he teaches Data Mining and Research methods. He has a strong interest in Machine Learning and is currently involved in several projects in this area, including classification methods for imbalanced data sets, dynamic credit scoring methods, contextual search techniques and mining high speed data streams.